\documentclass[10pt,twocolumn]{article}

\usepackage[utf8]{inputenc}
\usepackage[T1]{fontenc}
\usepackage[letterpaper,top=0.68in,bottom=0.71in,left=0.65in,right=0.65in,columnsep=0.22in]{geometry}
\usepackage{amsmath,amssymb,amsfonts}
\usepackage{booktabs}
\usepackage{array}
\usepackage{graphicx}
\usepackage{enumitem}
\usepackage{microtype}
\usepackage[hidelinks]{hyperref}

\setlist{nosep, leftmargin=1.4em}
\newcommand{\code}[1]{\texttt{#1}}
\newcommand\blfootnote[1]{%
  \begingroup
  \begin{NoHyper}%
  \renewcommand\thefootnote{}\footnote{#1}%
  \addtocounter{footnote}{-1}%
  \end{NoHyper}%
  \endgroup
}
\makeatletter
\renewcommand{\@seccntformat}[1]{%
  \ifcsname format@#1\endcsname
    \csname format@#1\endcsname
  \else
    \csname the#1\endcsname\quad
  \fi
}
\newcommand{\format@section}{}
\makeatother

\begin{document}

\title{\textbf{\texttt{blaris}: Windowless Random-Access LZSS for Constrained Microcontrollers}}

\author{%
  Ihor Lukianov\\
  \small\textit{Faculty of Mathematics and Computer Science, Jagiellonian University, Krak\'{o}w}\\
  \small\texttt{ihor.lukianov@student.uj.edu.pl}%
}

\date{}

\maketitle

\blfootnote{This work has been submitted to the IEEE for possible publication. Copyright may be transferred without notice, after which this version may no longer be accessible.}

\begin{abstract}
\noindent
Embedded LZ77 decompressors (such as \texttt{heatshrink} and \texttt{tamp})
require an in-RAM sliding window that couples memory usage to compression ratio
and incurs high latency on random-access reads. We present \texttt{blaris}, a
lightweight, windowless LZSS decompressor for execute-in-place (XIP)
random access on constrained microcontrollers. By partitioning tokens into
bidirectionally traversable streams with power-of-two checkpoints,
\texttt{blaris} resolves back-references directly in the compressed payload,
achieving compression ratios comparable to high-window \texttt{heatshrink} and
\texttt{tamp} configurations without allocating a RAM window. Targeting
compress-on-host, decompress-on-device workflows, \texttt{blaris} assumes
trusted build-time payloads and omits runtime error and bounds checks. On
RV32IMC, it requires just 548--784\,B of Flash and 104--112\,B of total RAM,
achieving orders-of-magnitude faster small-slice random reads than
ratio-matched, block-indexed \texttt{heatshrink} and \texttt{tamp}.
\end{abstract}

\section{Motivations}
On resource-constrained microcontrollers with only a few kilobytes of Flash and
RAM, every byte counts. Static assets such as UI bitmaps, fonts, device trees,
and configuration tables often consume large amounts of Flash even though they
are accessed infrequently.

Compressing these assets in Flash saves space, but general-purpose decompressors
like Zstandard~\cite{zstd} require over 100\,KiB of memory, and byte-aligned
codecs like LZ4 require multi-kilobyte blocks and windows to achieve competitive
ratios. Bit-level embedded alternatives such as
\code{heatshrink}~\cite{heatshrink}, \code{tamp}~\cite{tamp}, and
\code{uzlib}~\cite{uzlib} use far less memory, but as LZ77~\cite{lz77} stream
decoders, they still require an in-RAM sliding window to resolve
back-references. Because the window size directly determines both RAM usage and
compression ratio, better compression always costs more RAM.

Sliding-window decoders are also inefficient for random access to small data
slices. To read a few bytes from the middle of a compressed block, the decoder
must allocate the full window and sequentially decompress all preceding data
from the start of the block. Random-access parsings such as
LZ-End~\cite{lzend,kempa2022} avoid a sliding window by forcing matches to end
at prior phrase boundaries, but are designed for large, repetitive text
indexes~\cite{navarro2021} using auxiliary phrase tables and rank/select
bitvectors rather than kilobyte-scale MCU budgets.

This paper introduces \code{blaris}, a windowless variant of LZSS~\cite{lzss}
that removes the sliding-window buffer entirely. By organizing the compressed
data into streams that can be traversed both forward and backward, \code{blaris}
uses a small, fixed amount of RAM regardless of the lookback range. We further
optimize random-access reads by adding compact state checkpoints, host-side
optimal parsing, and reference-chain flattening.

\section{Windowless LZSS}
Standard LZSS compresses data into a sequence of tokens, where each token is
either a raw literal run or a back-reference match $(d, m)$ that copies $m$
bytes from a distance of $d$ bytes earlier. Conventional decoders keep recently
decompressed bytes in a RAM sliding window so that matches can be copied
immediately.

\code{blaris} eliminates the RAM window by resolving each byte directly within
the compressed payload. To decode a byte at an uncompressed offset $X$, the
decoder finds the token covering $X$:
\begin{enumerate}
    \item If the token is a \textbf{literal run}, the byte is read directly
        from the \code{literals} stream and emitted.
    \item If the token is a \textbf{match} with distance $d$, the target byte is
        identical to the byte at offset $X - d$. The decoder updates $X
        \leftarrow X - d$ and walks backward through the compressed tokens to
        find the token covering the new offset.
\end{enumerate}
This process repeats, following the chain of matches backward until it lands on
a literal run. In practice, rather than resolving one byte at a time, the
decoder tracks the remaining contiguous span $w$ (initialized to the requested
slice length) and narrows $w \leftarrow \min(w, P + \text{length} - X)$ at each
hop, copying all $w$ resolved bytes from the final literal run at once. While
walking token chains takes more CPU cycles than copying from a RAM buffer, it
completely removes the sliding window.

To make walking forward and backward across tokens fast and deterministic,
\code{blaris} splits the compressed payload into three separate streams:
\begin{itemize}
    \item \textbf{controls} (1 byte per token): values $0..15$ encode a literal
        run of $1..16$ bytes; values $16..255$ encode a match, packing all 240
        combinations of match length ($2..16$ bytes) and distance bit-width
        ($1..16$ bits).
    \item \textbf{literals} (1 byte per literal): stores the raw literal bytes
        contiguously.
    \item \textbf{distance\_bits} (bit-packed stream): stores the variable-width
        distance bits for each match token.
\end{itemize}

Because every control token is exactly 1 byte and specifies the exact number of
literal bytes or distance bits it consumes, the decoder can step both forward
and backward across all three streams in lockstep. The decoder tracks a 4-tuple
state $(P, C, L, D)$, representing the uncompressed byte position $P$, the
\code{controls} index $C$, the \code{literals} index $L$, and the bit offset $D$
in \code{distance\_bits}:
\begin{itemize}
    \item \textbf{Step forward}: read \code{controls[$C$]}. For a literal run of
        length $k$, add $k$ to $P$ and $L$. For a match of length $m$ with $b$
        distance bits, add $m$ to $P$ and $b$ to $D$. Increment $C$ by $1$.
    \item \textbf{Step backward}: decrement $C$ by $1$, read
        \code{controls[$C$]}, and subtract the corresponding $k$ or $(m, b)$
        values from $P$, $L$, and $D$.
\end{itemize}

With up to 16 distance bits per match, \code{blaris} addresses a 64\,KiB
lookback range. To reach 128\,KiB or 256\,KiB, match distances can optionally be
aligned to powers of two, trading a small loss in unaligned matches for a wider
lookback reach.

\section{Optimizations}

\subsection{Checkpoints}
Walking backward token by token uses constant $\mathcal{O}(1)$ RAM, but stepping
across thousands of tokens for a single back-reference is slow.
To keep lookups fast, we add \emph{checkpoints}: precomputed
snapshots of the decoder state $(P, C, L, D)$ stored in the payload header at
compression time.

Storing four raw 32-bit integers per checkpoint would cost 16 bytes each,
hurting the compression ratio on small files. To shrink this overhead, we place
checkpoints along a regular power-of-two grid so that the $i$-th checkpoint
starts at the first token at or after grid boundary $i \cdot
2^{\text{SPREAD\_LOG}}$:
\begin{equation}
    i \cdot 2^{\text{SPREAD\_LOG}} \le P_i < i \cdot 2^{\text{SPREAD\_LOG}} + 16.
\end{equation}
Because the maximum token length is 16 bytes, position $P_i$ is always within
$0..15$ bytes of the grid boundary. Its upper bits are already known from the
checkpoint index $i$, so we only need to store a 4-bit remainder for $P_i$.
Similarly, we split the bit offset $D$ into a byte index $(D \gg 3)$ and a 3-bit
remainder $(D \mathbin{\&} 7)$. Packing the 4-bit $P$ remainder and 3-bit $D$
remainder into one metadata byte uses just 7 bits. Depending on the chosen
offset width (2, 3, or 4 bytes per stream index), each checkpoint takes only 7,
10, or 13 bytes. To prevent unaligned memory reads on strict architectures,
checkpoints are split into two parallel arrays:
\begin{itemize}
    \item \textbf{metas} (1 byte per checkpoint): packs the 4-bit $P$ remainder
        and 3-bit $D$ bit-offset.
    \item \textbf{offsets} ($3 \times \{2,3,4\}$ bytes per checkpoint): stores
        the byte-aligned $(C, L, D \gg 3)$ indices.
\end{itemize}

Because checkpoints sit on a power-of-two grid, the decoder can find the
checkpoint nearest to any target offset $X$ in $\mathcal{O}(1)$ time using a
single bit-shift ($X \gg \text{SPREAD\_LOG}$). If the checkpoint's $+0..15$ byte
remainder places it slightly past $X$, the decoder simply steps backward by one
token. When following a match to $X' = X - d$, the decoder jumps to the nearest
checkpoint whenever doing so saves at least 32 bytes of linear token walking,
reducing each hop to a few token steps.

\subsection{Optimal Parser \& Chain Flattening}
\code{blaris} treats compression ratio as its primary goal. Because compression
runs offline on the host, encoding speed does not matter. We therefore adapt the
dynamic programming optimal parser from \code{lzmpo}~\cite{lzmpo} (similar to
shortest-path LZ parsing in LZMA~\cite{lzma}) to find the exact sequence of
literal runs and matches with the smallest total bit cost.

The parser first builds a single-byte hash chain linking each position $i$ to
preceding positions $j < i$ that share the same byte value. It then scans
forward from $i = 0$ to $N$. At each position $i$, it walks the hash chain
across the 64--256\,KiB lookback range to populate a local table
$\code{sheet}[2..16]$, where $\code{sheet}[k]$ stores the cheapest valid match
distance for an exact match of length $k$. Next, a downward pass from $k = 15$
down to $2$ replaces $\code{sheet}[k]$ with $\code{sheet}[k+1]$ whenever a
longer match is cheaper in bits than a shorter one. Using this table, the parser
evaluates all match transitions ($i \to i + k$) and literal-run transitions ($i
\to i + c$ for $c \in 1..16$) to update the minimum bit cost to reach each
subsequent position. Once the forward pass reaches $N$, the parser backtracks
from $N$ to $0$ along the cheapest transitions to emit the final tokens.

However, optimizing purely for size often creates deep chains of back-references,
where one match points to an earlier match, which points to another match, and
so on. Decoding a single byte can then require following dozens of match hops
before finally reaching a literal, slowing down random reads even with
checkpoints.

To shorten reference chains without sacrificing our size-first design, we add
optional \emph{chain flattening} (\code{f1}) as a post-processing pass on the
host. If a match token $X$ falls completely inside an earlier match token $Y$
that points to $Z$, chain flattening rewires $X$ to point directly to $Z$ (as
long as the combined distance fits within the 16-bit limit), skipping $Y$
entirely. Comparing \code{blaris\_o2\_a0\_s8\_f0} against \code{f1} (2-byte
stream offsets, $2^0=1$-byte alignment, $2^8=256$-byte checkpoint spread,
flattening off vs.\ on; Table~\ref{tab:jumps}), flattening cuts average
reference jumps by 14\%--51\% across 17 of the 21 corpus files at a 0.5\%--2.2\%
compression ratio cost (for example, halving average jumps on
\code{zcu102\_arm.dtb} from 45.6 to 22.4 at a 1.9\% ratio cost). The remaining 4
files are incompressible (\code{cp.html.gz}, $0\%$) or firmware binaries
dominated by short local matches (\code{ch32v30x\_blink.bin}, $-0.6\%$) or
multi-kilobyte zero-padding (\code{rv32i\_add.bin} and
\code{rv32i\_fence\_i.bin}, $0\%$). On the two zero-padded binaries, the optimal
parser encodes zero blocks as self-overlapping run-length matches
(\code{length=16, distance=1}), causing the decoder to step backward by $d=1$
byte per hop across the entire zero run and preventing whole-token chain
flattening.

\begin{table}[!t]
\centering
\caption{Impact of Chain Flattening (\code{o2\_a0\_s8\_f0} vs.\ \code{f1}) on
Reference Jumps (9 Representative Files; Full 21-File Table in Repository)}
\label{tab:jumps}
\scriptsize
\renewcommand{\arraystretch}{1.0}
\setlength{\tabcolsep}{1.3pt}
\begin{tabular}{@{}l r r r r r r r@{}}
\toprule
& \multicolumn{3}{c}{\textbf{Unflattened (\code{f0})}} & \multicolumn{3}{c}{\textbf{Flattened (\code{f1})}} & \\
\cmidrule(lr){2-4} \cmidrule(lr){5-7}
\textbf{Dataset} & \textbf{Ratio} & \textbf{Avg} & \textbf{Max} & \textbf{$\Delta$Ratio} & \textbf{Avg} & \textbf{Max} & \textbf{$\Delta$Avg} \\
\midrule
\code{FreeRTOSConfig.h}          &  35.5\% &   13.7 &  143 & $+1.6\%$ &   11.2 &  140 & $-18.2\%$ \\
\code{ch32v003\_boot.bin}        &  95.4\% &    0.6 &    7 & $+0.5\%$ &    0.6 &    7 & $-14.1\%$ \\
\code{fvp\_base\_gicv3.dtb}      &  33.5\% &   12.9 &   53 & $+1.5\%$ &    7.4 &   29 & $-42.5\%$ \\
\code{grammar.lsp}               &  43.8\% &    6.2 &   23 & $+1.1\%$ &    4.8 &   19 & $-21.9\%$ \\
\code{hifive-unleashed.dts}      &  51.3\% &    3.7 &   19 & $+1.0\%$ &    2.6 &   17 & $-30.2\%$ \\
\code{rv32i\_add.bin}            &  18.5\% & 3421.4 & 7699 & $+0.1\%$ & 3421.3 & 7696 & $0.0\%$   \\
\code{speed\_7578.csv}           &  23.3\% &   71.3 &  254 & $+1.1\%$ &   50.1 &  175 & $-29.8\%$ \\
\code{zcu102\_arm.dtb}           &  25.9\% &   45.6 &  226 & $+1.9\%$ &   22.4 &   81 & $-50.8\%$ \\
\code{zynq7000s.dtb}             &  34.5\% &   16.9 &   62 & $+1.5\%$ &   12.8 &   49 & $-23.9\%$ \\
\bottomrule
\end{tabular}
\end{table}

\section{Benchmarking Setup}
We evaluated \code{blaris} against \code{heatshrink} and \code{tamp} across a
21-file corpus covering microcontroller firmware binaries, C source code, web
assets (HTML/JSON), compiled Devicetree blobs (\code{.dtb}), and sensor CSV
logs. All benchmarks were compiled with Clang (\code{-Os}) for 32-bit RISC-V
(RV32IMC) and run in \code{rv32emu}~\cite{rv32emu} at 1 cycle per instruction
with zero-wait-state memory.

For each file and codec configuration, we tested 9 slice sizes from $2^0$ (1\,B)
to $2^8$ (256\,B). At each size, we decompressed 1,000 pseudorandom unaligned
offsets and measured execution cycles via the RISC-V \code{rdcycle64} counter.
We measured exact peak stack usage by filling the stack with \code{0xA5} canary
bytes before execution and scanning for the deepest modified byte afterward.

All decompressors were wrapped in a common random-access API
(\code{init}, \code{slice}, and \code{reset}). Because \code{heatshrink} and
\code{tamp} only decode sequentially, we added a block-indexing wrapper that
splits files into independently compressed blocks ($2^4$ to $2^{14}$\,B) with a
header offset table, allowing them to seek directly to the target block.
Table~\ref{tab:memory} summarizes total Flash (\code{.text} + \code{.rodata})
and RAM (peak stack + \code{.bss} + \code{.data}) across all 206 tested
configurations (108 \code{blaris}, 74 \code{heatshrink}, 24 \code{tamp}).
Because \code{blaris} targets trusted build-time payloads, it omits runtime
bounds and error checks to keep code size minimal.

Finally, to check whether SPI Flash wait states on real hardware
disproportionately penalize \code{blaris}'s backward token walking, we tested a
representative balanced configuration of each codec across the corpus on a
physical ESP32-C3FN4 microcontroller in direct-boot XIP mode, mapping both
decompressor code and the compressed payload to external quad-SPI Flash without
copying either to SRAM.

\begin{table}[!t]
\centering
\caption{Decompressor Memory Footprints (RV32IMC, Including Adapter)}
\label{tab:memory}
\footnotesize
\renewcommand{\arraystretch}{1.15}
\setlength{\tabcolsep}{2.4pt}
\begin{tabular}{@{}l r r r r@{}}
\toprule
\textbf{Codec} & \textbf{Flash} & \textbf{Stack} & \textbf{Static RAM} & \textbf{Total RAM} \\
\midrule
\code{blaris}     &   548--784\,B   &      76\,B &    28--36\,B      &   104--112\,B \\
\code{heatshrink} & 1,300--1,464\,B & 140--144\,B &    66--16,450\,B &   206--16,594\,B \\
\code{tamp}       & 2,238--3,238\,B & 228--248\,B &   292--16,424\,B &   520--16,672\,B \\
\bottomrule
\end{tabular}
\end{table}

\section{Results}
\begin{figure}[!t]
\centering
\makebox[\columnwidth][c]{\includegraphics[width=1.05\columnwidth]{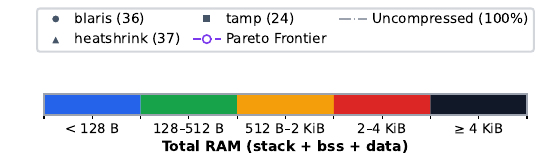}}\\[1pt]
\makebox[\columnwidth][c]{\includegraphics[width=1.05\columnwidth]{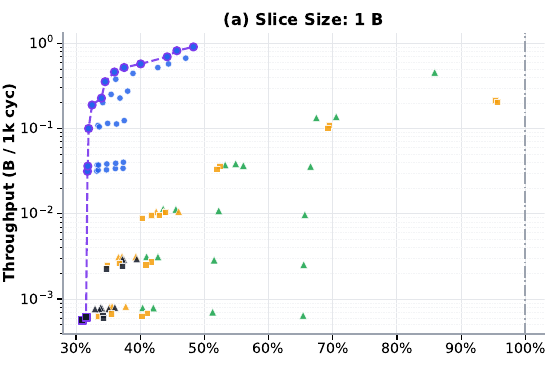}}\\[0.5pt]
\makebox[\columnwidth][c]{\includegraphics[width=1.05\columnwidth]{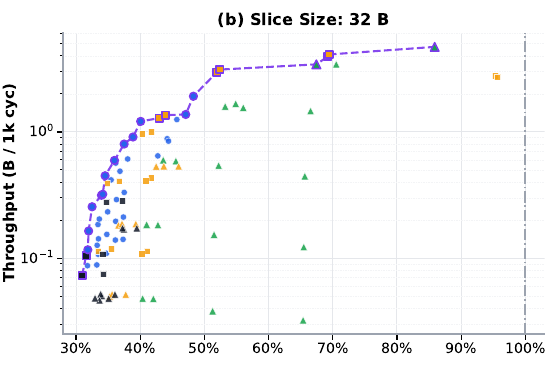}}\\[0.5pt]
\makebox[\columnwidth][c]{\includegraphics[width=1.05\columnwidth]{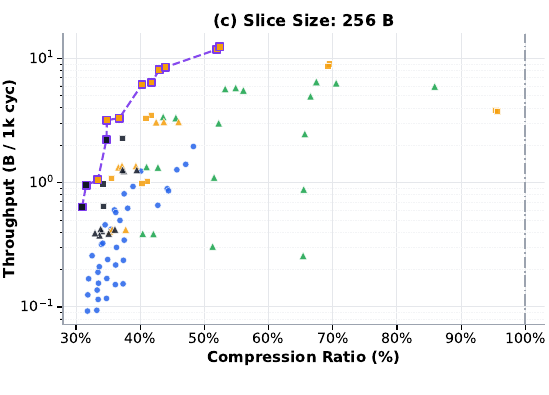}}
\caption{Decompression throughput vs.\ compression ratio on \code{zynq7000s.dtb}
(15,583\,B) at slice sizes of (a) 1\,B, (b) 32\,B, and (c) 256\,B across 97
configurations (\code{blaris\_o3}/\code{o4} and \code{heatshrink\_i32} omitted
for clarity).}
\label{fig:zynq_pareto}
\end{figure}

Figure~\ref{fig:zynq_pareto} shows the Pareto frontier of decompression speed
versus compression ratio at slice sizes of 1\,B, 32\,B, and 256\,B on a
representative Devicetree file, \code{zynq7000s.dtb} (15.2\,KiB).

On 1\,B random reads (Fig.~\ref{fig:zynq_pareto}a), \code{blaris} holds 12 of
the 14 Pareto-optimal points and runs over two orders of magnitude faster than
ratio-matched \code{heatshrink} and \code{tamp} configurations. Because smaller
independent blocks degrade compression ratio, matching \code{blaris}'s $\sim 36\%$
ratio requires configuring \code{heatshrink} and \code{tamp} with a 1\,KiB
window (\code{w10}) and 16\,KiB blocks (\code{b14}). At this ratio on
\code{zynq7000s.dtb}, \code{blaris\_o2\_a0\_s8\_f1} takes 2,173 cycles/byte
(0.46\,B/kcyc) at a 35.99\% ratio using 112\,B of RAM, outperforming
\code{heatshrink\_b14\_i16\_w10\_l6} ($562\times$ slower, 35.63\% ratio,
1,214\,B RAM) and \code{tamp\_b14\_w10\_e0} ($685\times$ slower, 35.54\% ratio,
1,288\,B RAM). Similarly, on \code{FreeRTOSConfig.h} (35.7\,KiB), \code{blaris}
(2,311 cycles/byte, 37.10\% ratio) achieves both a better ratio and $598\times$
and $641\times$ higher speed than the same \code{heatshrink} (43.83\%) and
\code{tamp} (41.31\%) configurations.

At 32\,B slices (Fig.~\ref{fig:zynq_pareto}b), \code{blaris} still holds 12 of
the 16 sub-50\%-ratio Pareto points (reaching 1.92\,B/kcyc) while using only
104--112\,B of total RAM. High-window \code{tamp} and \code{heatshrink}
configurations join the faster end of the frontier, but require $5\times$--$40\times$
more RAM (520\,B to $>4$\,KiB) for \code{tamp} and $2\times$--$4\times$ more
(206--462\,B) for \code{heatshrink}.

At 256\,B slices (Fig.~\ref{fig:zynq_pareto}c), block-streaming codecs overtake
\code{blaris}. Because \code{blaris} resolves matches without caching prior
output in RAM, its cost grows linearly with slice length, whereas streaming
decoders pay a one-time block-seek cost and then emit bytes cheaply. This places
the crossover around 64--128\,B slices: below this size, \code{blaris} offers
superior speed and RAM efficiency; above it, streaming codecs are faster when
sufficient RAM is available.

Comparing \code{rv32emu} against physical cached-XIP ESP32-C3FN4 silicon on one
balanced configuration per codec (\code{blaris\_o2\_a0\_s8\_f1} across 19 files,
excluding the two zero-padded \code{rv32i\_add.bin} and
\code{rv32i\_fence\_i.bin} binaries due to their $\mathcal{O}(N)$ RLE jump
depth, and \code{heatshrink\_b10\_i16\_w8\_l4} and \code{tamp\_b10\_w8\_e1}
across all 21 files) shows that SPI Flash wait states do not disproportionately
penalize \code{blaris} relative to the baselines. Overall hardware cycle counts
increase by 29.3\% for \code{blaris}, $33.5\%$
for \code{heatshrink}, and $23.6\%$ for
\code{tamp}. On 1\,B reads, \code{blaris} sees a $1.51\times$ hardware overhead,
likely because cold SPI Flash cache-line fills make up a larger share of its
short execution time; for slices $\ge 4$\,B, the overhead stabilizes at
$1.30\times$--$1.33\times$ ($1.30\times$--$1.31\times$ for $\ge 8$\,B).

\section{Conclusion}
We presented \code{blaris}, a windowless LZSS decompressor designed for deeply
constrained microcontrollers where RAM budgets are too tight for
\code{heatshrink} or \code{tamp}. By resolving back-references directly in the
compressed payload via bidirectional streams and compact power-of-two
checkpoints, \code{blaris} achieves compression ratios comparable to high-window
\code{heatshrink} and \code{tamp} configurations (which require $>1.2$\,KiB of
RAM) while using only 548--784\,B of Flash and 104--112\,B of total RAM. For
random access to small slices, \code{blaris} is up to orders of magnitude faster
than ratio-matched, block-indexed streaming decoders, though streaming codecs
become faster on larger sequential reads ($\ge 128$\,B) when speed is the
priority and sufficient RAM is available.

The complete \code{blaris} implementation, host compressors, and reproducible
benchmark suite are available as open-source software at
\url{https://github.com/lis05/blaris}.

\end{document}